

\font\titolino=cmbx10
\font\tsnorm=cmr10
\font\tscors=cmti10

\font\tscorsp=cmti9
\magnification=1200
\hsize=148truemm
\hoffset=10truemm
\parskip 3truemm plus 1truemm minus 1truemm
\parindent 8truemm
\newcount\notenumber

\def\scc{\scriptscriptstyle}
\def\PR{{\tscors Phys. Rev. }}
\def\NP{{\tscors Nucl. Phys. }}
\def\PL{{\tscors Phys. Lett. }}
\def\NC{{\tscors Nuovo Cimento }}

\def\JMP{{\tscors J. Math. Phys. }}
\def\IJMP{{\tscors Int. J. Math. Phys  }}
\def\CQG{{\tscors Class. Quantum Grav. }}
\def\note{\advance\notenumber by 1 \footnote{$^{\the\notenumber}$}}
\def\ref#1{\medskip\everypar={\hangindent 2\parindent}#1}
\def\beginref{\begingroup
\bigskip
\leftline{\titolino References.}
\nobreak\noindent}
\def\endref{\par\endgroup}
\def\beginsection #1. #2.
{\bigskip
\leftline{\titolino #1. #2.}
\nobreak\noindent}
\def\beginappendix #1.
{\bigskip
\leftline{\titolino Appendix #1.}
\nobreak\noindent}
\def\beginack
{\bigskip
\leftline{\titolino Acknowledgments}
\nobreak\noindent}

\def\S{\hbox{Schr\"o\-din\-ger }}
\def\ra{\rightarrow}

\nopagenumbers
\rightline{DFTT 23/93}
\rightline{ Ref. SISSA 64/93/A}
\rightline{April 29, 1993}
\rightline{Revised version}
\vskip 15truemm
\centerline{\titolino ON A QUANTUM UNIVERSE FILLED}
\bigskip
\centerline{\titolino WITH YANG - MILLS RADIATION}
\vskip 10truemm
\centerline{\tsnorm Marco Cavagli\`a$^{(a),(c)}$ and
Vittorio de Alfaro$^{(b),(c)}$}
\bigskip
\centerline{$^{(a)}$\tscorsp Sissa - Int. School for Advanced Studies
Trieste, Italy}
\bigskip
\centerline{$^{(b)}$\tscorsp Dipartimento di Fisica
Teorica dell'Universit\`a di Torino}
\bigskip
\centerline{$^{(c)}$\tscorsp INFN, Sezione di Torino, Italy}
\bigskip
\vskip 10truemm
\centerline{\tsnorm ABSTRACT}
\begingroup\tsnorm\noindent
We investigate the properties of a quantum Robertson - Walker universe
described by the Wheeler -- DeWitt equation. The universe is filled with
a quantum Yang -- Mills uniform field. This is then a quantum mini copy
of the standard model of our universe. We discuss the interpretation of
the Wheeler -- DeWitt wave function using the correspondence principle
to connect $\vert\psi\vert^2$ for large quantum numbers to the classical
probability for a radiation dominated universe. This can be done in any
temporal gauge. The correspondence principle determines the
Schr\"odinger representation of the momentum associated to the
gravitational degree of freedom. We also discuss the measure in the
mini--superspace needed to ensure invariance of the quantum description
under change of the temporal gauge. Finally, we examine the behaviour of
$\vert\psi\vert^2$ in inflationary conditions.
\vfill
\hrule
\noindent
Mail Address:
\hfill\break
Dipartimento di Fisica Teorica
\hfill\break
Via Giuria 1, I-10125 Torino
\hfill\break
Electronic mail:VAXTO::VDA or VDA@TO.INFN.IT
\endgroup
\vfill
\eject
\footline{\hfill\folio\hfill}
\pageno=1
\beginsection 1. Introduction.
We may believe that the universe at Planck size for the scale factor is
quantistic, but not too much, meaning that an underlying field structure
(strings or any sort of quantum field treatment) is not predominant;
thus a quantum treatment of a few degrees of freedom could be
sufficient. Then it is tempting to investigate the properties of the
Wheeler - DeWitt (WDW) equation [1,2] for a mini universe (a single
degree of freedom, the scale factor) filled by Yang - Mills (YM)
quantized radiation. This is then a quantum version of a radiation
dominated RW universe (we shall discuss only in the last section the
presence of an effective cosmological constant).

In a reasonable closed quantum universe, when the system is confined to
about a Planck length $L_p$ both the YM field and the gravitational
degree of freedom have small quantum numbers since the two are
connected. In order to get a Standard Model RW universe (confined, if it
is the case, to at least some 10$^{60}$ $L_p$), enormously large
excitation quantum numbers are required, which makes a quantum
description irrelevant and at the same time shows once more the
unnaturalness of the conditions for the present universe (in absence of
inflation of course) as seen from the quantum point of view.

However, this leads us to use the correspondence principle between
classical probability and  $\vert\psi\vert^2$ having imposed square
integrability conditions on the wave functions. If we allow the \S
representation  of the momentum associated to the scale factor $a$ to
depend on the gauge chosen for the classical time, then
$\vert\psi\vert^2$ satisfies the correspondence principle in any
temporal gauge. Then the wave function depends on the time gauge. We
will see how to define the quantum mechanical measure so that amplitudes
are gauge independent.

Finally we discuss briefly the behaviour of the wave function in
inflationary condition. Of course  a YM field is of no help in producing
inflation, that requires the usual ingredient, some effective
cosmological constant whose effect is to stretch enormously the
extension of the wave function depressing its amplitude. After inflation
and reheating, the radiation has no connection with the original
solution and the description of the universe is classical.
\beginsection 2. Classical preliminaries.
We write the gravitational action as
$$A_g=-\int d^4x\sqrt{-g}\bigl(R+2\Lambda\bigr)\eqno(2.1)$$
The signature of spacetime is $(+,-,-,-,)$; definitions as in Landau -
Lifshitz. We have put $16\pi G=~L_p^2~=~M_p^{-2}=1$ thus measuring all
dimensional quantities in these units.

We take the Robertson - Walker metric of topology $R\times S_3$ where
$S_3$ is the three- sphere or the euclidean flat space ($k$=1,0).
$$ds^2=N^2(\theta)d\theta^2-a^2(\theta)\omega^p\otimes\omega^p
\eqno(2.2)$$
$\omega^p$ are the 1-forms invariant under translations in space  and
$N(\theta)$ is the lapse function. The cosmic time $t$ corresponds to
$N=1$ and the conformal time $\tau$ to $N=a$. The $\omega^p$'s satisfy
the SU(2) Maurer-Cartan structure equation:
$$d\omega^p={k\over 2}\epsilon_{pqr}\omega^q\wedge\omega^r\eqno(2.3)$$
and thus (2.2) has the SU(2)$_L$$\times$SU(2)$_R$ group of isometries.

The action space density is
$$S_g=\int d\theta ~ L_g\eqno(2.4)$$
(We do not imply by this that the action is uniform in the 3-D space
with local topology $R_3$, only that we consider a region in which eqs.
(2.2,3) hold over a domain of the extension of order of $L_p$. This is
enough if inflation does the rest. See for instance [3].) $L_g$ is given
by
$$L_g=6\biggl(-{a\dot a^2\over N}+kNa\biggr)-2\Lambda Na^3\eqno(2.5)$$
$L_g$ has been obtained from (2.1) by integration by parts in time and
neglecting total derivatives. This is enough if we are just interested
in the classical equations of motion.

Let us now review briefly how to introduce a YM field configuration with
the same SU(2)$\times$SU(2) symmetry as the metric.  We shall use a YM
group SU(2) for simplicity (the general case has been investigated in
ref. [4]). It could be different in different regions of size $L_p$; this
choice has little to do with what will be found after filling the
universe after inflation when reheating will have created large
excitation numbers democratically for the whole set of YM fields.

The most general YM potential invariant under SU(2)$\times$SU(2) depends
essentially only on time.
%

The YM action density is
$$A_{YM}={1\over 2}\int F\wedge{}^*F\eqno(2.6)$$
where (we have set = 1 the gauge coupling constant) the field strength
2-form $F=dA+A\wedge A$ is written is terms of the 1-form potential
$$A=A^a_\mu(x){i\over 2}\sigma_a dx^\mu\eqno(2.7)$$
The most general form of the SU(2)$\times$SU(2) invariant field, written
in the vierbein cotangent space, is
$$A={i\over 2}\xi(\theta)\sigma_p\omega^p\eqno(2.8)$$
With the definition (2.8) $A$ is evidently left- invariant; it is also
right- invariant up to a gauge transformation [5,6]

It is straightforward to find the field strength $F$:
$$F={i\over 2}\sigma_p\dot\xi d\theta \wedge \omega^p~ +
{}~{i\over 4}\sigma_r\epsilon_{rpq}\xi(k-\xi) \omega^q \wedge \omega^r
\eqno (2.9)$$
Thus the physical space action density of the YM field is
$$S_{YM}=\int d\theta~ {3\over 2}
\bigl(N^{-1}a\dot\xi^2-Na^{-1}\xi^2(1-\xi)^2\bigr)
\eqno(2.10)$$
We give also the form of the energy - momentum tensor:
$$\eqalign{&T_{\theta \theta}={{3}\over{2}}N^2\bigl
(N^{-2}a^{-2}\dot\xi^{2}
+a^{-4}\xi^2(1-\xi)^2\bigr)\cr
&T_{ij}=-{1\over 2}\bigl(N^{-2}a^{-2}\dot\xi^{2}
+a^{-4}\xi^2(1-\xi)^2\bigr)g_{ij}\cr}\eqno(2.11)$$

To establish the WDW equation for the mini universe [1,7-10] let us
introduce the conjugate momenta
$$\eqalign{p_a&=-12{a\dot a\over N}\cr
p_\xi&=3N^{-1}a\dot\xi\cr}\eqno(2.12)$$
The Hamiltonian to be used in the WDW equation is
$$H={1\over 12a}\Biggl[-\biggl({1\over 2}p^2_a+V_a(a)\biggr)+
4\biggl({1\over 2}p^2_{\xi}+V_\xi(\xi)\biggr)\Biggr]\eqno(2.13)$$
where
$$\eqalign{V_a(a)&=72\biggl(ka^2-{\Lambda\over 3}a^4\biggr)\cr
V_\xi(\xi)&={9\over 2}\xi^2(1-\xi)^2\cr}\eqno(2.14)$$
The classical Friedmann - Einstein (FE) equation of motion is
$$H=0\eqno(2.15)$$
{}From the YM equation follows [5]
$$\bigl(N^{-1}a\dot\xi\bigr)^2+\xi^2(1-\xi)^2=K^2\eqno(2.16)$$
where $K$ is independent of $\theta$. Using (2.16), eq. (2.13) becomes
$${a^2\dot a^2\over N^2}+\biggl(ka^2-{\Lambda\over 3}a^4\biggr)={1\over
4}K^2\eqno(2.17)$$
In the cosmic gauge $N=1$ this reads ($\dot a \equiv da/dt$)
$${\dot a^2\over a^2}+{k\over a^2}={\Lambda\over 3}+{1\over 4}{K^2\over
a^4}\eqno(2.18)$$
We see that, as due to radiation, the energy density scales as $a^{-4}$.
The meaning of $K^2$ is apparent:
$$K^2={2\over 3}\rho(a=1)\eqno(2.19)$$
$K^2$ is the energy density of the radiation in a RW universe with scale
factor $a$ of one Planck length. Now suppose for a  moment we take our
present classical universe and run it back by the Standard Model, no
inflation, up to the time at which $a\simeq L_p$ ($\leq~10^{-61} a_0$).
Now, in the Standard Model the Planck density, $\rho=1$, is reached when
$a\simeq 10^{-31} a_0~\simeq 10^{30}L_p$. Then, going further to $a=
L_p$ the density must be sort of $10^{120}\rho_p$. $K^2$ would be a very
strange number indeed, in order to obtain our present universe by the
sole power of the Standard Model + YM radiation. This is only too well
known, it is just the old observation of mismatch between size and
density. We shall now see the quantum counterpart.
\beginsection 3. The WDW equation.
The WDW equation for a YM filled RW universe has been first discussed
in [11]. We will emphasize some different aspects, mainly the
correspondence principle and the connection between eigenvalues of the
gravitational and the YM sectors.

The WDW equation is given by
$$H\Psi(a,\xi)=0\eqno(3.1)$$
where $H$ is given by (2.13) and we deal now with quantum operators:
$$\hfill\bigl[a,p_a\bigr]=i;\hbox to15truemm{}
\bigl[\xi,p_\xi\bigr]=i\eqno(3.2)$$
In the quantum version of the Hamiltonian (2.13) there is an ordering
problem between the term $1/a$ and $p_a^2$. This may be connected to the
\S re\-pre\-sen\-ta\-tion of the momentum through the correspondence
principle and will be discussed later. For the moment we simply drop
that overall factor $1/a$ from the Schr\"o\-din\-ger equa\-tion
$H\psi=0$.

It is a beautiful property of the radiation field [11] that in the
quantum equation there is no direct coupling between the YM field and
the gravitational degree of freedom (similar to the case of a conformal
scalar field discussed in [7], but physically much more interesting).
This is the quantum form of the request that the density scales like
$a^{-4}$. The two fields are in reality essentially coupled, since it is
the presence of the YM part that allows the gravitational degree of
freedom to have a quantum solution.  The state  of the quantum mini
universe is determined by the quantum configuration of the YM field: the
presence of a non vanishing eigenvalue of the YM sector supports a non
trivial gravitational configuration.

Thus, the WDW equation separates. We write
$$\Psi(a,\xi)=\psi(a)\eta(\xi)\eqno(3.3)$$
and look for eigenvalues of the YM part:
$$\biggl[{1\over 2}p^2_\xi+V_\xi(\xi)\biggr]\eta_n(\xi)=E_{n}^{YM}\eta_n(\xi)
\eqno(3.4)$$
Now about the boundary conditions for the YM wave function. It is
natural to ask that the wave function tends to zero for large $|\xi|$.
Then eigenvalues are quantized. $n$ counts the number of oscillations.
It is easy to derive, and can be controlled by the WKB expansion [12],
that the asymptotic behaviour of $E_n$ for large $n$ is $E_n\sim
n^{4/3}$.

Now let us turn to the gravitational degree of freedom. We drop the
cosmological term and discuss the case $k=1$ (see also [13] where the
WDW equation in presence of classical matter is discussed too as a
quantum bound state). We have
$$\biggl[{1\over 2}p^2_a+V_a(a)~\biggr]\psi_n(a)=E^g_n\psi_n(a)
\eqno(3.5)$$
The gravitational degree of freedom is given by a harmonic oscillator. It
is natural to set the boundary condition at $a\rightarrow \infty$ by
asking the square integrability of the wave function (for this suggestion
see also [14]). With this criterion one obtains the correspondence with
the classical gravitational motion for large oscillator quantum numbers
as we shall see. About the condition at $a=0$, if we ask that
$\psi\rightarrow 0$, then $p_{\xi}$ (and thus $H_{YM}$) is hermitean
(this does not mean however that $p_{\xi}$ is observable, as its
eigenfunctions do not fulfil the boundary condition). Let us adhere to
these boundary conditions. Then the quantum numbers $n_g$ are odd. We
have
$$E^g_n=12\biggl({1\over 2}+n_g\biggr)\eqno(3.6)$$
The eigenvalues of the two eigenequations are connected:
$$E^g_n=4E^{YM}_n\eqno(3.7)$$
In this equation two constants are actually present: the Newton constant
and the YM coupling. It may look that this equality could hold only for
a few states; however in order to have a mini universe one just needs it
to be valid for one state. The passage to macroscopic configurations
requires inflation.

If for a moment we use the numbers quoted before for our classical
universe (run back to $a \sim L_p$ by the Standard Model without
inflation), we would need $E^g_n\sim n_g\sim 10^{120}$, and $n_{YM}\sim
10^{90}$ (see [1]). With such artificial quantum numbers both degrees of
freedom should be described classically. This displays how unnatural is
our universe without inflation in the frame of the WDW equation for YM
fields.

In this context it is interesting to notice that one should be able to
recover some information about the quantum wave function and its meaning
using the principle of correspondence between classical and quantum
systems for large $n$ (as suggested originally in [1]).
\beginsection 4. The wave function and the correspondence principle.
Let us examine with some care the gravitational wave function $\psi(a)$.
We are discussing  a closed RW universe filled with radiation. The point
is, in eq. (3.5) which is the \S representation of $p_a$? is it the
naive one, namely
$$p_a~\ra~-i~d/da~~?\eqno(4.1)$$
It is tempting to interpret $\vert \psi\vert^2$ as the probability
density for the value $a$ for the scale factor [8], to be compared
through the correspondence principle to the classical probability
density. This implies that in some way $\vert \psi\vert^2$ depends on
$N(a)$ (we consider $N$ as function of the mini--superspace variable
$a$, as for instance in the case of the conformal gauge). We will see
how that can be implemented. If one takes this attitude, then the
correspondence principle is the guidance to solve this problem. The idea
is that the \S representation of $p_a$ is to be determined by the
request that for large quantum numbers $\vert \psi\vert^2$ approaches,
in average, the classical probability distribution of the physical
quantity $a$ for an ensemble of trajectories. Let us first discuss this
point with some care. The WDW equation is independent of time. This is
because time is just a degree of freedom in the universe, being a
measure of correlation of positions of  physical objects [1,15]. Now in
our WDW equation the universe has been reduced to just two degrees of
freedom; we have no hands nor clocks in the hamiltonian (2.13). So, to
compare the quantum system to a classical one we may consider the
classical motion for an ensemble forgetting their starting time. We are
taking a snapshot at the scale factor $a$ in the classical system and do
not know how much time has lapsed since its classical beginning. We only
ask for the probability (density) of finding the classical system at a
certain $a$. This is inversely proportional to the speed of $a$ in time.
Thus (probability not normalized)
$$P_{cl}^{\theta}~=~{1\over{da/d\theta}}\eqno(4.2)$$
%
%
{}From the classical equation of motion (2.17) for $k=1$, $\Lambda=0$ and
$K^2/4\equiv a_M^2$ we get
$$P_{cl}^{\theta}~=~{a\over N(a)}~
{1\over\sqrt{a_M^2-a^2}}\eqno(4.3)$$
where $N(a)=1$ for the cosmic time and $N(a)=a$ for the conformal time.
Now if this probability has to be compared to $\vert \psi\vert^2$ by the
correspondence principle, we must admit that $\psi$ is not invariant
under change of the lapse function $N(a)$, in spite of the fact that the
WDW equation is independent of $N$ (see also [14]). A way out is the
suggestion that the rule for the representation of $p_a$ is
$$p_a~\ra~\sqrt{{a\over N(a)}}~\biggl(-i{d\over da}\biggr)
\sqrt{{N(a)\over a}}\eqno(4.4)$$
Indeed, with $\psi_n=\sqrt {a/N(a)}~ \chi_n$ the eigenvalue equation
reads
$$\biggl(-{1\over 2}{d^2\over da^2}+V_a(a)\biggr)\chi_{n}(a)
=E_{n}^g~ \chi_{n}(a)\eqno(4.5)$$
Then the correspondence principle works properly:
$$\Sigma_{av}~\vert\psi_n(a)\vert^2~\ra~ {a\over N(a)}~
{1\over\sqrt{a_M^2-a^2}}\eqno(4.6)$$
coincident with the classical probability (4.3). The representation
(4.4) gives the required result. Thus the correspondence principle
suggests that the Schr\"o\-din\-ger re\-pre\-sen\-ta\-tion of $p_a$, and
thus of $\psi(a)$, depends on the gauge chosen for the time.

Let us check how these ideas work for the case of a flat RW universe,
$k=0$. In this case, in the cosmic gauge the classical equation of
motion (2.18) is
$$a\dot a=\hbox{constant}\eqno(4.7)$$
Thus
$$P_{cl}^t~\propto~a\eqno(4.8)$$
Now the corresponding quantum equation is
$$-{1\over 2}{d^2\over da^2}~{\psi(a)\over\sqrt a}~ =
{}~E{\psi(a)\over\sqrt a}\eqno(4.9)$$
The solution is $\psi(a)=~C\sqrt a e^{\pm ia\sqrt{2E}}$ whose
probability distribution gives back (4.8) (we have nothing new to say
about the problem of the boundary conditions in this case).

It is interesting to observe that in the conformal gauge ($N(a)=a$) the
representation for $p_a$ is just the naive one. The conformal gauge is
privileged in this respect. We will see the reason.

The choice (4.4) implies that the integrals representing scalar products
in the Hilbert space have the measure
$$(\Psi_1, O(p_a,a;p_\xi,\xi)\Psi_2)~=~\int \Psi_1^*
O(p_a,a;p_\xi,\xi)\Psi_2 {N(a)\over a} da~ d\xi \eqno(4.10)$$
First of all notice that any matrix element of the form (4.10) is lapse
independent. Indeed we have
$$(\psi_1,O(p_a,a)\psi_2)~=~ \int \chi_1^*O(-id/da,a)\chi_2~ da
\eqno(4.11)$$
Nor the equation for the $\chi$'s nor the boundary conditions depend on
$N(a)$ (in particular $H$ is hermitian if $\chi(0)=0$). One sees how the
measure compensates the dependence of the wave function on the lapse
function required by the correspondence principle.

Thus, the matrix elements that have the fundamental quantum role are
given by (4.10). In particular, the invariant probability with the usual
quantum properties is $dP=\vert \psi\vert^2 (N(a)/a) da$. It is curious
but true that the squared modulus of the wave function obeys the
correspondence principle as we discussed, but is not a true probability
density since (as the corresponding classical density) it is not
invariant under changes of the time gauge.

Let us observe that our procedure agrees with the Halliwell discussion
[16] of the rescaling of the lapse function. The difference is that we
allow for the lapse dependence of the representation of the wave
function and this guarantees the invariance of matrix elements under
changes of $N(a)$. The conformal time gauge has a special role since
when $N=a$ the measure in (4.10) is the elementary one.

Let us note that in our simplified quantum frame a quantity that could
be interpreted a posteriori, at least for large quantum numbers, as a
substitute for a classical measure of ``time interval'' is
$$\Delta \theta~=\int_{a_1}^{a_2}\vert\psi\vert^2~da\eqno(4.12)$$
as suggested by the classical interpretation. It is hard to see how this
could be generalized.
\beginsection 5. Inflation.
Let us discuss the behaviour of the wave function and control the
correspondence principle in an inflationary region. For these limited
purposes in our quantum mini universe we stimulate inflation through the
introduction of an effective cosmological parameter $\Lambda_e(a)$ which
is constant for small $a$ and soon or later tends to 0. This expedient
may simulate both the case of chaotic inflation and of the quenched
$\Lambda$ cases.

Let us turn our attention to the classical equation of motion (2.17) for
the gravitational degree of freedom. (we use now the cosmic gauge for
definiteness and $k=0$ for simplicity, to avoid discussing tunnelling):
$$a^2\dot a^2-{\Lambda\over 3}a^4={E\over 72}\eqno(5.1)$$
Thus
$$P_{cl}^t={1\over\dot a}={12a\over\bigl[2(E_n+24\Lambda a^4)\bigr]^{1/2}}
\eqno(5.2)$$
Notice that for large $a$ $P_{cl}\sim 1/a$. The physical interpretation
is that during the de Sitter phase the scale factor $a$ changes
exponentially in time, $\dot a\sim a$, thus the time the classical
system spends in a neighbourhood of any given value is inversely
porportional to the scale factor.

The scale at which this behaviour sets in is of course
$(E_n/24\Lambda)^{1/4}$. We see that the effect of the YM eigenvalue
$E_n$ is to delay inflation: the larger is the energy, the higher the
value of $a$ at which inflation starts.

Now we turn to the WDW equation (in the cosmic gauge, and $k=0$):
$$\biggl(-{d^2\over da^2}-48\Lambda a^4\biggr)\chi_n(a)~
=~2E_n~\chi_n(a)\eqno(5.3)$$
The WKB solution solution (we are far from turning points; if
$\Lambda>0$ there are none for $E_n>0$) is given by (we are very
interested in the pre-exponential factor and not in the outgoing or
incoming boundary conditions)
$$\chi_n(a)={{C}\over{\sqrt{p_n(a)}}}e^{\pm i\int^a_0p_{\scc n}(a)da}
\eqno(5.4)$$
where
$$p_n(a)=\bigl[2(E_n+24\Lambda a^4)\bigr]^{1/2}\eqno(5.5)$$
Remembering now that in this gauge $\psi_n(a)=\sqrt a\chi_n(a)$ we
obtain for the quantum probability distribution the same formula as for
the classical one. In particular for sufficiently large $a$ the WKB wave
function has pure de Sitter behaviour:
$$\vert\psi(a)\vert^2\sim{1\over a}\eqno(5.6)$$
The inflationary stretching of the wave function from the Planck domain
into large domains depresses its amplitude.

There is no way in this quantum context to tell how big is the size of
the de Sitter region in $a$. Adhering to the chaotic inflation scenario,
$\Lambda\sim V(\phi)$, this depends on the parameters of the potential.
One knows that one needs a huge factor, certainly larger than $10^{31}$.
This is of course unnatural in terms of the Planck wave function, but,
that is the way it goes with inflation. 
requires an inflationary model 
classical behaviour of $a$ in time. Quantum mechanics alone cannot help
us particularly. Note that the integral of the squared wave function
over the de Sitter region is of the order of the classical inflation
time:
$$\int_{a_1}^{a_2}\vert\psi\vert^2~da\sim \Lambda^{-1/2} \ln (a_2/a_1)
\eqno(5.7)$$
To end with the case of a cosmological constant, let us remark that if
$\Lambda$ is negative and $k$=0 the hamiltonian of the two degrees of
freedom has the same form (apart from the sign). The condition that both
are e.g. in the lowest quantum eigenstate requires that $\vert \Lambda
\vert$ and the squared gauge coupling constant are equal. This is the
quantum analogue of the classical solutions discussed long ago [17].
\beginack
We are very grateful to T. Regge for interesting discussions on
various questions raised by the subject of this paper. Correspondence
with L.J. Garay and O. Bertolami is gratefully acknowledged. We thank
also M. Gasperini, M. Pauri and G. Veneziano.
\bigskip
\centerline{\bf{* * *}}
\bigskip
After the submission of this paper we have realized that our suggestion
(4.12) coincides with  the ``probabilistic time'' extensively explored
by M.A. Castagnino and F. Lombardo, \PR {\bf D48}, 1722 (Aug. 15, 1993).
\vfill\eject
\beginref

\ref [1] B. DeWitt, \PR {\bf 160}, 1113 (1967).

\ref [2] J.A. Wheeler, in: {\tscors Battelle Rencontres} eds. C. DeWitt
and J.A. Wheeler, Benjamin, 1968.

\ref [3] A.D. Linde, in: {\tscors 300 Years of Gravitation}, eds. S.W.
Hawking and W. Israel, Cambridge University Press, 1987.

\ref [4] O. Bertolami, J.M. Mour\~ao, R.F. Picken and I.P. Volobujev,
\IJMP {\bf A23}, 4149 (1991).

\ref [5] M. Henneaux, \JMP {\bf 23}, 830 (1982).

\ref [6] D.V. Galt'sov and M.S. Volkov, \PL {B256}, 17 (1991).

\ref [7] S.W. Hawking, \NP {\bf B239}, 257 (1984).

\ref [8] S.W. Hawking and D. Page, \NP {\bf B264}, 185 (1986).

\ref [9] A. Vilenkin, \PR {\bf D37}, 888 (1988).

\ref [10] A. Strominger, \NP {\bf B319}, 722, (1989).

\ref [11] O. Bertolami and J.M. Mour\~ao, \CQG {\bf 8}, 1271 (1991).

\ref [12] C. Rossetti, {\tscors Rivista} \NC {\bf 12}, N. 8, (1989).

\ref [13] J. H. Kung, Harvard--Smithsonian Center for Astrophysics,
Preprint: HEP-TH/9302016 (1993).

\ref [14] L.J. Garay, Preprint: GR-QC/9306002.

\ref [15] T. Banks, \NP {\bf B49}, 332 (1985).

\ref [16] J.J. Halliwell, in: {\tscors Conceptual Problems of Quantum
Gravity}, eds. by A. Ashtekar and J. Stachel, Birkh\"auser, Boston, 1991.

\ref [17] V. de Alfaro, S. Fubini and G. Furlan, \NC {\bf 50} N. 4, 523,
(1979).

\endref
\bye